\documentclass[prb,superscriptaddress,showpacs,floatfix,tightenlines,twocolumn]{revtex4}
\usepackage{amssymb}
\usepackage{amsmath}
\usepackage{graphicx}

\begin{document}

\title{Dynamical matrix of bidimensional electron crystals}
\author{R. C\^{o}t\'{e}}
\affiliation{D\'{e}partement de physique and RQMP, Universit\'{e} de Sherbrooke,
Sherbrooke, Qu\'{e}bec, Canada, J1K 2R1}
\author{M.-A. Lemonde}
\affiliation{D\'{e}partement de physique and RQMP, Universit\'{e} de Sherbrooke,
Sherbrooke, Qu\'{e}bec, Canada, J1K 2R1}
\author{C. B. Doiron}
\affiliation{Department of Physics, University of Basel, Klingelbergstrasse 82, CH-4056
Basel, Switzerland}
\author{A. M. Ettouhami}
\affiliation{Department of Physics, University of Toronto, 60 St. George St., Toronto,
Ontario, Canada, M5S 1A7}

\begin{abstract}
In a quantizing magnetic field, the two-dimensional electron (2DEG) gas has
a rich phase diagram with broken translational symmetry phases such as
Wigner, bubble, and stripe crystals. In this paper, we derive a method to
get the dynamical matrix of these crystals from a calculation of the density
response function performed in the Generalized Random Phase Approximation
(GRPA). We discuss the validity of our method by comparing the dynamical
matrix calculated from the GRPA\ with that obtained from standard elasticity
theory with the elastic coefficients obtained from a calculation of the
deformation energy of the crystal.
\end{abstract}

\date{\today }
\maketitle

\section{Introduction}

Theoretical calculations show that, in the presence of a perpendicular
magnetic field, a two-dimensional electron gas (2DEG) should crystallize
below a filling factor $\nu \sim 1/6.5$\cite{lamgirvin}. Several
experimental groups have reported transport measurements indicative of this
electron crystallization when the filling factor of the lowest Landau level
is decreased below $\nu =1/5.$ These measurements include the observation of
a strong increase in the diagonal resistivity $\rho _{xx},$ non-linear $I-V$
characteristics, and broadband noise. All these observations have been
interpreted as the pinning and sliding of a Wigner crystal (WC)\cite%
{reviewexperimentwc}. Moreover, microwave absorption experiments\cite{ye}
have also detected a resonance in the real part of the longitudinal
conductivity, $\sigma _{xx}\left( \omega \right)$, that has been attributed to
the pinning mode of a disordered Wigner crystal. The vanishing of the
pinning mode resonance at some critical temperature $T_{m}\left( \nu \right) 
$ has been used to derive the phase diagram of the crystal\cite{yongchen} in
the quantum regime where the kinetic energy is frozen by the quantizing
magnetic field. Similar microwave absorption experiments also showed a
pinning resonance at higher filling factors close to $\nu =1,2,3$ where the
formation of a Wigner solid is expected in very clean samples\cite%
{chen1,lewis1,lewis2}. Finally, in Landau levels of index $N>1$, a study of
the evolution of the pinning mode with filling factor reveals several
transitions of the 2DEG ground state from a Wigner crystal at low $\nu $ to
bubble crystals with increasing number of electrons per lattice site as
$\nu $ is increased, and into a modulated stripe state (or anisotropic Wigner
crystal) near half filling\cite{chen2}.

In earlier works\cite{cotemethode,ettouhami1}, some of us have studied several crystalline states of the
2DEG using a combination of Hartree-Fock (HFA) and generalized random-phase
approximations (GRPA). In these works, the
energy and order parameters of the crystal were calculated in a
self-consistent HFA while the collective excitations were derived from the
poles of density response functions computed in the GRPA. This microscopic
approach (HFA + GRPA) works well at zero temperature but is difficult to
generalize to consider finite temperature effects or to include quantum
fluctuations beyond the GRPA. Finite-temperature or quantum fluctuations
effects (not already included in the GRPA), are most easily computed by
writing down an elastic action for the system. For a crystaline solid, this
requires the knowledge of the dynamical matrix (DM) or, equivalently, of
the elastic coefficients of the solid.

A direct way to obtain these elastic coefficients is
to compute the energy required for various static deformations of the
crystal. Using elasticity theory, each deformation energy $\Delta E_{i}$ can
be written in the form $\Delta E_{i}=\frac{1}{2}C_{i}u_{0}^{2}$ where $u_{0}$
is a parameter characterizing the amplitude of the deformation and $C_{i}$
is generally a combination of elastic constants. In the limit $%
u_{0}\rightarrow 0$, one can obtain the elastic coefficients by computing
the deformation energy of one or more static deformations and using the
known symmetry relations between the elastic constants. Alternatively, one
can obtain a DM from the GRPA density response function much more directly
without the need to compute the elastic coefficients \cite{cote}.

In this paper, we compare the DM obtained from these two methods
(deformation energy and GRPA) in order to find the range of validity as well
as the limitations of the GPRA approach. We first consider the simple case of an
isotropic (triangular) Wigner crystal before tackling the more complex anisotropic
Wigner crystal \cite{ettouhami1} or stripe phase that occurs near half-filling in the higher
Landau levels. We show that although the GRPA method gives a good description
of the qualitative behavior of the DM as a function of filling factor, its quantitative
predictions must be used with caution. As we show below, an averaging procedure must be
applied to the method in order to obtain a DM in the GRPA that compares favorably with the one
obtained by computing the deformation energy.

Our paper is organized as follows. In Section II, we define the elastic
constants needed to build an elastic model for the Wigner and stripe
crystals. We then explain, in Section III, how these elastic constants can be derived by
computing the deformation energy of the crystals in the HFA. In section IV, we
summarize the GRPA method of obtaining the dynamical matrix. Our numerical
results for the WC are discussed in Section V and those for the stripe
crystal in Section VI. Section VI contains our conclusions.

\section{Elastic constants and dynamical matrix}

We describe the elastic deformation of a crystal state by a displacement
field $\mathbf{u}\left( \mathbf{R}\right) $ defined on each lattice site $%
\mathbf{R}$. The Fourier transform of this operator is given by:
\begin{equation}
\mathbf{u}\left( \mathbf{k}\right) =\frac{1}{\sqrt{N_{s}}}\sum_{\mathbf{R}%
}e^{-i\mathbf{k}\cdot \mathbf{R}}\mathbf{u}\left( \mathbf{R}\right) ,
\end{equation}%
where $N_{s}$ is the number of lattice sites. In two dimensions, the general
expression for the deformation energy of a crystal requires the use of $6$ elastic
coefficients $c_{ij}$ and is given, in the continuum limit, by the following expression\cite{landau}: 
\begin{align}
\Delta E& =\frac{1}{2}\int d\mathbf{r}\left[
c_{11}e_{x,x}^{2}+4c_{66}e_{x,y}^{2}+4c_{62}e_{x,y}e_{yy}\,\right]
\label{m4} \\
& +\frac{1}{2}\int d\mathbf{r}\left[
2c_{12}e_{x,x}e_{y,y}+4c_{16}e_{x,x}e_{x,y}+c_{22}e_{y,y}^{2}\,\right] , 
\notag
\end{align}
where 
\begin{equation}
e_{\alpha ,\beta }\left( \mathbf{r}\right) =\frac{1}{2}\left( \frac{\partial
u_{\alpha }\left( \mathbf{r}\right) }{\partial r_{\beta }}+\frac{\partial
u_{\beta }\left( \mathbf{r}\right) }{\partial r_{\alpha }}\right)
\end{equation}
is the symmetric strain tensor.

The Wigner and bubble crystals have a triangular lattice structure for which
the following equation holds: 
\begin{equation}
c_{11}=c_{22}=2c_{66}+c_{12}.  \label{triangulaire}
\end{equation}
For such a triangular structure, the elastic energy in the long-wavelength limit can be
written in a form that contains only two elastic coefficients, namely:
\begin{eqnarray}
\Delta E &=&\frac{1}{2}\int d\mathbf{r}\left[ c_{12}\left(
e_{x,x}^{2}+e_{y,y}^{2}+2e_{x,x}e_{y,y}\right) \right.  
\notag \\
&+&\left. 2c_{66}\left( e_{x,x}^{2}+e_{y,y}^{2}+2e_{x,y}^{2}\right) \right].
\label{m6}
\end{eqnarray}

The anisotropic stripe state can be seen either as a centered rectangular
lattice with two electrons per unit cell or as a rhombic lattice with one
electron per unit cell with reflection symmetry in both $x$ and $y$ axis.
The deformation energy is given by 
\begin{align}
\Delta E& =\frac{1}{2}\int d\mathbf{r}\left[
c_{11}e_{x,x}^{2}+4c_{66}e_{x,y}^{2}\,\right] 
\notag\\
& +\frac{1}{2}\int d\mathbf{r}\left[
2c_{12}e_{x,x}e_{y,y}+c_{22}e_{y,y}^{2}\,\right]. 
\end{align}
In this paper, we assume that the stripes are aligned along the $y$ axis.

The above formulation of elasticity theory assumes short-range forces only.
For the electronic crystals that we consider, these forces are of coulombic
origin i.e. the hamiltonian of the crystal contains only the Coulomb
interaction between electrons and the kinetic energy which is frozen by the
quantizing magnetic field. Both the direct (Hartree) and exchange (Fock)
terms are considered by the Hartree-Fock approximation as we explain in the
next section. To take into account in the elasticity theory the long-range
part of the Coulomb interaction present in a crystal of electrons, it is
necessary to add to $\Delta E$ the deformation energy $\Delta E_{C}$ given by%
\begin{eqnarray}
\Delta E_{C} &=&\frac{e^{2}}{2}\int d\mathbf{r}\int d\mathbf{r}^{\prime }%
\frac{\delta n\left( \mathbf{r}\right) \delta n\left( \mathbf{r}^{\prime
}\right) }{\kappa \left\vert \mathbf{r-r}^{\prime }\right\vert } \\
&=&\frac{\pi e^{2}}{S}\sum_{\mathbf{q}}\frac{\delta n\left( \mathbf{q}%
\right) \delta n\left( \mathbf{-q}\right) }{\kappa q},  \notag
\end{eqnarray}
where $S$ is the area of the crystal, $\delta n\left( \mathbf{q}\right)
=\int d\mathbf{r}e^{-i\mathbf{q}\cdot \mathbf{r}}\delta n\left( \mathbf{r}
\right) $ is the Fourier transform of the change in the electronic density
and $\kappa $ is the dielectric constant of the host semiconductor. We
consider the positive background of ionized donors as homogeneous and inert
so that no linear term in $\delta n\left( \mathbf{r}\right) $ is introduced
by the Coulomb interaction.

To define a dynamical matrix, we assume that the crystal can be viewed as a
lattice of electrons with static form factor $h\left( \mathbf{r}\right) $ on
each crystal site (with the normalisation $\int d\mathbf{r}h\left( \mathbf{r}%
\right) =1$). The time-dependent density can then be written as 
\begin{equation}
n\left( \mathbf{r},t\right) =\sum_{\mathbf{R}}h\left( \mathbf{r}-\mathbf{R}-%
\mathbf{u}\left( \mathbf{R},t\right) \right) ,  \label{m13}
\end{equation}%
and, to first order in the displacement field, we have for a density
fluctuation

\begin{equation}
\delta n\left( \mathbf{k+G},t\right) =-ih\left( \mathbf{k+G}\right) \sqrt{%
N_{s}}\left( \mathbf{k}+\mathbf{G}\right) \cdot \mathbf{u}\left( \mathbf{k}%
,t\right) ,  \label{m7}
\end{equation}%
where $\mathbf{G}$ is a reciprocal lattice vector and $\mathbf{k}$ a vector
in the first Brillouin zone of the crystal. It follows that we can write the
Coulomb energy as%
\begin{equation}
\Delta E_{C}=\pi n_{0}e^{2}\sum_{\mathbf{q}}\frac{\left\vert h\left( \mathbf{%
q}\right) \mathbf{q}\cdot \mathbf{u}\left( \mathbf{q}\right) \right\vert ^{2}%
}{\kappa q},  \label{defo}
\end{equation}%
where $n_{0}=N_{s}/S$ is the average electronic density.

We pause at this point to remark that the form factor $h\left( q\right) \sim
e^{-q^{2}\ell ^{2}/2}$ (with $\ell =\sqrt{\hslash c/eB}$ the magnetic
length, $B$ being the applied magnetic field) in Eq. (\ref{defo}) 
renders the summation over the wavectors rapidly
convergent. Our Hartree-Fock calculation of the ground-state energy of the
electronic crystals as well as our GRPA calculation of the dynamical matrix
also involve summations over reciprocal lattice vectors $\mathbf{G}$ of some
functions weighted by $h\left( G\right) $. If the magnetic field is not too
strong, we can perform these summations directly. There is no need to use
Ewald's summation technique as is the case if one works with a crystal of
point electrons. Of course, as the filling factor $\nu \rightarrow 0$, the
magnetic length $\ell \rightarrow 0$ so the electrons behave more and more
like point particles and the convergence is lost. In all cases that we
consider, the summations involved are rapidly convergent because we restrict
ourselves to filling factors $\nu =2\pi n_0\ell ^{2}\gtrsim 0.1$ where $\ell/a_{0}$ 
is sufficiently large for $e^{-G^{2}\ell ^{2}/2}$ to be small ($a_{0}$ being the lattice constant). 
The cutoff in $G$ is choosen so that the
summations are evaluated with the required degree of accuracy.

The total deformation energy, which we now write as $\Delta E_{T},$ now
includes the long-range Coulomb interaction and can be written in the form:
\begin{equation}
\Delta E_{T}=\frac{1}{2}\sum_{\mathbf{k}}u_{\alpha }\left( \mathbf{k}\right)
D_{\alpha ,\beta }\left( \mathbf{k}\right) u_{\beta }\left( -\mathbf{k}
\right) ,  
\label{m18}
\end{equation}
where we have introduced the dynamical matrix:
\begin{equation}
D_{\alpha ,\beta }\left( \mathbf{k}\right) =\frac{\partial ^{2}\Delta E_{T}}{%
\partial u_{\alpha }\left( \mathbf{k}\right) \partial u_{\beta }\left( -%
\mathbf{k}\right) }.
\end{equation}%
For the triangular lattice, a comparison of Eqs. (\ref{m18}) and  (\ref{m6}) 
gives the dynamical matrix (to order $k^{2}$) as: 
\begin{subequations}
\label{m32}
\begin{eqnarray}
D_{x,x}\left( \mathbf{k}\right) &=&n_{0}^{-1}\left[ \left( \widetilde{c}%
_{12}\left( k\right) +2c_{66}\right) k_{x}^{2}+c_{66}k_{y}^{2}\right] , \\
D_{x,y}\left( \mathbf{k}\right) &=&n_{0}^{-1}\left( \widetilde{c}_{12}\left(
k\right) +c_{66}\right) k_{x}k_{y}, \\
D_{y,y}\left( \mathbf{k}\right) &=&n_{0}^{-1}\left[ \left( \widetilde{c}%
_{12}\left( k\right) +2c_{66}\right) k_{y}^{2}+c_{66}k_{x}^{2}\right] .
\end{eqnarray}
\end{subequations}
The long-range Coulomb interaction renders the elastic coefficient $c_{12}$
(but not the shear modulus $c_{66}$) nonlocal, so that $c_{12}$ contains a
diverging term $\sim 1/k$. We shall write:
\begin{equation}
\widetilde{c}_{12}\left( k\right) =\frac{2\pi n_{0}^{2}e^{2}}{\kappa k}%
+c_{12},  \label{a55}
\end{equation}
where $c_{12}$ is the weakly dispersive part of the elastic coefficient,
and where the plasmonic (first) term on the {\em rhs} is due to the long-range 
nature of the Coulomb interaction.

For the stripe state, Eq. (\ref{triangulaire}) is no longer valid.
In addition, all three elastic coefficients $c_{11},c_{12},c_{22}$
become nonlocal. We have in this case:
\begin{subequations}
\begin{eqnarray}
D_{x,x}\left( \mathbf{k}\right) &=&n_{0}^{-1}\left[ \widetilde{c}_{11}
\left( k\right) k_{x}^{2}+c_{66}k_{y}^{2}\right]
\nonumber\\
&+&n_{0}^{-1}Kk_{y}^{4}, \\
D_{x,y}\left( \mathbf{k}\right) &=&n_{0}^{-1}\left( \widetilde{c}_{12}\left(
k\right) +c_{66}\right) k_{x}k_{y}, \\
D_{y,y}\left( \mathbf{k}\right) &=&n_{0}^{-1}\left[ 
\widetilde{c}_{22}\left( k\right) k_{y}^{2}+c_{66}k_{x}^{2}\right] ,
\end{eqnarray}
\end{subequations}
where $\widetilde{c}_{ij}=\frac{2\pi n_{0}^{2}e^{2}}{\kappa k}+c_{ij}$, with 
$i,j=1,2$; and where we added to $D_{x,x}\left( 
\mathbf{k}\right) $ a term $Kk_{y}^{4}$ in order to take into account the
bending rigidity of the stripes which, due to the small value of the shear modulus $c_{66}$ 
in these systems, is quantitatively important 
over a sizeable region of the Brillouin zone\cite{ettouhami1}. 
Using the fact that $n_{0}=\nu/2\pi \ell ^{2}$, we finally obtain:
\begin{equation}
\widetilde{c}_{ij}=\left( \frac{e^{2}}{\kappa \ell }\right) 
\frac{\nu }{k\ell }n_{0}+c_{ij}.
\end{equation}
We now want to discuss how one can evaluate the non-dispersive part 
$c_{ij}$ of the elastic coefficients. This will be the subject of
the follwing section.

\section{Calculation of the elastic coefficients in the Hartree-Fock
approximation}

In the Hartree-Fock approximation, a crystalline phase is described by the
Fourier components $\left\langle n\left( \mathbf{G}\right) \right\rangle $
of the average electronic density, where $\mathbf{G}$ is a reciprocal lattice vector. In
the strong magnetic field limit where the Hilbert space is restricted to one
Landau level, it is more convenient to work with the \textquotedblleft
guiding-center density\textquotedblright\ $\left\langle \rho \left( \mathbf{G}\right) \right\rangle $ 
which is related to $\left\langle n\left( \mathbf{G}\right) \right\rangle $ by: 
\begin{equation}
\left\langle n\left( \mathbf{G}\right) \right\rangle =N_{\varphi
}F_{N}\left( \mathbf{G}\right) \left\langle \rho \left( \mathbf{G}\right)
\right\rangle ,  
\label{nphi}
\end{equation}
\ where $N_{\varphi }$ is the Landau-level degeneracy and
\begin{equation}
F_{N}\left( \mathbf{G}\right) =e^{-G^{2}\ell ^{2}/4}L_{N}^{0}\left( \frac{%
G^{2}\ell ^{2}}{2}\right)  \label{nphi2}
\end{equation}%
is the form factor of an electron in Landau level $N$ ($L_{N}^{0}\left(x\right)$ 
being a generalized Laguerre polynomial). The magnetic field $\mathbf{B}
=B\widehat{\mathbf{z}}$ is perpendicular to the 2DEG.

The Hartree-Fock energy per electron in the partially filled Landau level is
given by\cite{cotemethode,ettouhami1}:
\begin{equation}
\frac{E}{N_{e}}=\frac{1}{2\nu }\sum_{\mathbf{G}}\left[ H\left( \mathbf{G}%
\right) \left( 1-\delta _{\mathbf{G},0}\right) -X\left( \mathbf{G}\right) %
\right] \left\vert \left\langle \rho \left( \mathbf{G}\right) \right\rangle
\right\vert ^{2},  \label{energiehf}
\end{equation}%
where the $\delta _{\mathbf{G},0}$ term in this equation accounts for the
neutralizing background of the ionized donors. The parameter $\nu
=N_{e}/N_{\varphi }$ is the filling factor of the partially filled level, and
we take all filled levels below $N$ to be inert. The Hartree and Fock
interactions in Landau level $N$ are defined by: 
\begin{subequations}
\begin{eqnarray}
H\left( \mathbf{q}\right) &=&\left( \frac{e^{2}}{\kappa \ell }\right) \frac{1%
}{q\ell }e^{\frac{-q^{2}\ell ^{2}}{2}}\left[ L_{N}^{0}\left( \frac{q^{2}\ell
^{2}}{2}\right) \right] ^{2}, \\
X\left( \mathbf{q}\right) &=&\left( \frac{e^{2}}{\kappa \ell }\right) \sqrt{2%
}\int_{0}^{\infty }dx\,e^{-x^{2}} \\
&&\times \left[ L_{N}^{0}\left( x^{2}\right) \right] ^{2}J_{0}\left( \sqrt{2}%
xq\ell \right) ,  \notag
\end{eqnarray}%
where $J_{0}\left( x\right) $ is the Bessel function of the first kind.

To compute the $\left\langle \rho \left( \mathbf{G}\right) \right\rangle
^{\prime }s$, we first write this quantity in second quantization and in the
Landau gauge $\mathbf{A}=\left( 0,Bx,0\right) $ as: 
\end{subequations}
\begin{equation}
\left\langle \rho (\mathbf{G})\right\rangle =\frac{1}{N_{\varphi }}%
\sum_{X}e^{-iG_{x}X+iG_{x}G_{y}\ell ^{2}/2}\ \left\langle c_{N,X}^{\dagger
}c_{N,X-G_{y}\ell ^{2}}\right\rangle .
\end{equation}%
The average values $\left\langle \rho (\mathbf{G})\right\rangle $ are
obtained by computing the single-particle Green's function
(here and in what follows, $T_\tau$ denotes the time ordering operator): 
\begin{equation}
G\left( X,X^{\prime },\tau \right) =-\left\langle T_{\tau} c_{N,X}\left( \tau
\right) c_{N,X^{\prime }}^{\dagger }\left( 0\right) \right\rangle ,
\end{equation}%
whose Fourier transform we define as: 
\begin{equation}
G\left( \mathbf{G,}\tau \right) =\frac{1}{N_{\phi }}\sum_{X,X^{\prime }}e^{-%
\frac{i}{2}G_{x}\left( X+X^{\prime }\right) }\delta _{X,X^{\prime
}-G_{y}\ell ^{2}}G\left( X,X^{\prime },\tau \right) ,
\end{equation}%
so that: 
\begin{equation}
\left\langle \rho \left( \mathbf{G}\right) \right\rangle =G\left( \mathbf{G,}%
\tau =0^{-}\right) .
\end{equation}

We use an iterative scheme to solve numerically\cite{cotemethode} the
Hartree-Fock equation of motion for $G\left( \mathbf{G,}\tau \right) .$ For
the undeformed lattice, we use the basis vectors: 
\begin{subequations}
\begin{eqnarray}
\mathbf{R}_{1} &=&a_{0}\eta \sin \left( \varphi \right) \widehat{\mathbf{x}}%
+a_{0}\eta \cos \left( \varphi \right) \widehat{\mathbf{y}}, \\
\mathbf{R}_{2} &=&a_{0}\widehat{\mathbf{y}},
\end{eqnarray}
\end{subequations}
where $\eta $ is the aspect ratio and $\varphi $ is the angle between the
two basis vectors. For the triangular lattice, $\eta =1$ and $\varphi =\pi/3 $. 
If we apply an elastic deformation $\mathbf{u}\left( \mathbf{r}\right) 
$ to the lattice, the new lattice vectors are given by $\mathbf{R}^{\prime
}=n\mathbf{R}_{1}+m\mathbf{R}_{2}+\mathbf{u}\left( \mathbf{r}\right) $
(where $n,m$ are integers). We can write this expression as 
$\mathbf{R}^{\prime }=n\mathbf{R}_{1}^{\prime }+m\mathbf{R}_{2}^{\prime }$ if we define
the new basis vectors as: 
\begin{subequations}
\begin{eqnarray}
\mathbf{R}_{1}^{\prime } &=&a_{0}^{\prime }\eta ^{\prime }\sin \left(
\varphi ^{\prime }\right) \widehat{\mathbf{x}}+a_{0}^{\prime }\eta ^{\prime
}\cos \left( \varphi ^{\prime }\right) \widehat{\mathbf{y}}, \\
\mathbf{R}_{2}^{\prime } &=&a_{0}^{\prime }\widehat{\mathbf{y}}.
\end{eqnarray}
\end{subequations}
The parameters $a_{0}^{\prime },\eta ^{\prime }$ and $\varphi ^{\prime }$
are functions of the original lattice and of the type of deformation
considered. The reciprocal lattice vectors of the deformed lattice are
easily computed once these parameters are known. Then, the cohesive energy $E(u_0)$
of the deformed lattice can be calculated using the deformed reciprocal
lattice vectors and Eq. (\ref{m7}). Under these circumstances, we find that
the deformation energy per electron is given by: 
\begin{equation}
f=\frac{E\left( u_{0}\right) }{N_{e}}-\frac{E\left( u_{0}=0\right) }{N_{e}}.
\label{deformation}
\end{equation}%
To find the elastic coefficients for the Wigner and stripe crystals, we need
to consider the following deformations (note that the magnetic field and the
number of electrons are kept fixed\cite{ettouhami1,ettouhami2}):

{\it (i)} A shear deformation with $u_{x}\left( \mathbf{r}\right) =u_{0}y$ and 
$u_{y}\left( \mathbf{r}\right) =0$: the strain tensors in this case are given by 
$e_{x,x}\left( \mathbf{r}\right) 
=e_{y,y}\left( \mathbf{r}\right)=0$, and $e_{x,y}\left( \mathbf{r}\right) =u_{0}/2$. 
The area of the system, $S$, is not changed by this
deformation and the elastic energy is given by $F_{shear}=\frac{1}{2}Sc_{66}u_{0}^{2}$.
It then follows that the shear modulus $c_{66}$ is given by: 
\begin{equation}
c_{66}=\lim_{u_{0}\rightarrow 0}n_{0}\frac{d^{2}f_{shear}}{du_{0}^{2}},
\label{m1}
\end{equation}%
where $f=F/N_{e}$ is the deformation energy per electron. The parameters of
the distorted lattice for this shear deformation are given by: 
\begin{subequations}
\begin{align}
&a_{0}^{\prime } = a_{0}, 
\\
&\eta ^{\prime } =\eta \sqrt{1+u_{0}\sin \left( \varphi \right) \cos \left(
\varphi \right) +u_{0}^{2}\sin ^{2}\left( \varphi \right) }, 
\\
&\sin \left( \varphi ^{\prime }\right) =\frac{\sin \left( \varphi \right) }{%
\sqrt{1+u_{0}\sin \left( \varphi \right) \cos \left( \varphi \right)
+u_{0}^{2}\sin ^{2}\left( \varphi \right) }}.
\end{align}
\end{subequations}

\medskip

{\it (ii)} A one-dimensional dilatation along $\widehat{\mathbf{x}}$, with 
$u_{x}\left( \mathbf{r}\right) =u_{0}x$ and $u_{y}\left( \mathbf{r}\right) =0$:
here, the strain tensors $e_{x,x}\left( \mathbf{r}\right) =u_{0},e_{y,y}\left( 
\mathbf{r}\right) =0,$ and $e_{x,y}\left( \mathbf{r}\right) =0$, and the new
area of the system is $S^{\prime }=\left( 1+u_{0}\right) S$ and $F_{dx}=
\frac{1}{2}Sc_{11}u_{0}^{2}$. It then follows that the compression constant $c_{11}$ is given by: 
\begin{equation}
c_{11}=\lim_{u_{0}\rightarrow 0}n_{0}\frac{d^{2}f_{dx}}{du_{0}^{2}},
\label{m2}
\end{equation}
while the parameters of the deformed lattice are given by: 
\begin{subequations}
\begin{align}
&a_{0}^{\prime } =a_{0}, 
\\
&\eta ^{\prime } =\eta \sqrt{1+\left( 2u_{0}+u_{0}^{2}\right) \sin^{2}\left( \varphi \right) }, 
\\
&\sin \left( \varphi ^{\prime }\right) =\frac{\left( 1+u_{0}\right) \sin
\left( \varphi \right) }{\sqrt{1+\left( 2u_{0}+u_{0}^{2}\right) \sin^{2}\left( \varphi \right) }}.
\end{align}
\end{subequations}
The surface of the deformed lattice is $S^{\prime }=a_{0}^{\prime 2}\eta^{\prime }\sin \left( \varphi ^{\prime }\right) =S\left( 1+u_{0}\right)$,
so that the filling factor is now given by $\nu^{\prime }=\nu /\left( 1+u_{0}\right)$.

\medskip

{\it (iii)} A one-dimensional dilatation along $\widehat{\mathbf{y}}$ with 
$u_{x}\left( \mathbf{r}\right) =0$ and $u_{y}\left( \mathbf{r}\right) =u_{0}y$:
now the strain tensors $e_{x,x}\left( \mathbf{r}\right) =0,e_{y,y}\left( 
\mathbf{r}\right) =u_{0}$ and $e_{x,y}\left( \mathbf{r}\right) =0$. The new
area of the system is $S^{\prime }=\left( 1+u_{0}\right) S$ and $F_{dy}=
\frac{1}{2}Sc_{22}u_{0}^{2}.$ The compression constant $c_{22}$ is therefore given
by: 
\begin{equation}
c_{22}=\lim_{u_{0}\rightarrow 0}n_{0}\frac{d^{2}f_{dy}}{du_{0}^{2}}.
\label{m3}
\end{equation}%
On the other hand, the parameters of the deformed lattice are given by: 
\begin{subequations}
\begin{align}
&a_{0}^{\prime } =\left( 1+u_{0}\right) a_{0}, 
\\
&\eta ^{\prime } =\frac{\eta }{\left( 1+u_{0}\right) }\sqrt{1+\left(
2u_{0}+u_{0}^{2}\right) \cos ^{2}\left( \varphi \right) }, 
\\
&\cos \left( \varphi ^{\prime }\right) =\frac{\left( 1+u_{0}\right) \cos
\left( \varphi \right) }{\sqrt{1+\left( 2u_{0}+u_{0}^{2}\right) \cos
^{2}\left( \varphi \right)}}.
\end{align}
\end{subequations}
The surface of the deformed lattice is $S^{\prime }=a_{0}^{\prime 2}
\eta^{\prime }\sin \left( \varphi ^{\prime }\right) =S\left( 1+u_{0}\right)$,
so that the filling factor $\nu ^{\prime }=\nu /\left( 1+u_{0}\right) .$

\medskip

{\it (iv)} A two-dimensional dilatation with $u_{x}\left( \mathbf{r}\right)
=u_{0}x$ and $u_{y}\left( \mathbf{r}\right) =u_{0}y$: now, the strain tensors 
$e_{x,x}\left( \mathbf{r}\right) = e_{y,y}\left( \mathbf{r}\right)
=u_{0} $ and $e_{x,y}\left( \mathbf{r}\right) =0$. The new area of the
system is $S^{\prime }=\left( 1+u_{0}\right) ^{2}S$ and 
$F_{dxy}=\frac{1}{2}S\left( c_{11}+2c_{12}+c_{22}\right) u_{0}^{2}.$ It follows that the
combination $c_{11}+2c_{12}+c_{22}$ is given by: 
\begin{equation}
c_{11}+2c_{12}+c_{22}=\lim_{u_{0}\rightarrow 0}n_{0}\frac{d^{2}f_{dxy}}{%
du_{0}^{2}}.  \label{m4p}
\end{equation}
For this case, there is no need to actually compute the energy of the
deformed lattice since we can extract $c_{11}+2c_{12}+c_{22}$ from the
Hartree-Fock energy $E/N_{e}$ given in Eq. (\ref{energiehf}) in the
following manner. The area per electron, $s$, in the deformed lattice is $%
s=\left( 1+u_{0}\right) ^{2}s_{0}$, so that ($s_0$ here is the area per
electron of the undeformed lattice):
\begin{equation}
c_{11}+2c_{12}+c_{22}=4s_{0}\left( \frac{d^{2}f_{dxy}}{ds^{2}}\right)_{s=s_{0}}.
\end{equation}
The change in $s$ causes a change in the filling factor, which is now given by: 
\begin{equation}
\nu ^{\prime }=\frac{\nu}{\left( 1+u_{0}\right)^2} = \frac{2\pi \ell^2}{s^{\prime }}.
\end{equation}
Writing the HF energy as $E/N_{e}=\left( \frac{e^{2}}{\kappa \ell }\right) A\left( \nu\right) $, 
we have the relation: 
\begin{equation}
c_{11}+2c_{12}+c_{22}=\frac{2}{\pi }\left( \frac{e^{2}}{\kappa \ell ^{3}}%
\right) \nu ^{2}\left[ \nu \frac{d^{2}A\left( \nu \right) }{d\nu ^{2}}+2%
\frac{dA\left( \nu \right) }{d\nu }\right] .  \label{m9}
\end{equation}
Note that the long-wavelength Coulomb term $\frac{2\pi n_{0}^{2}e^{2}}{\kappa k}$ must
be added to $c_{11},c_{12}$ and $c_{22}$ that we compute in order to get 
$\widetilde{c}_{11},\widetilde{c}_{12}$ and $\widetilde{c}_{22}$.

Fig. 1 shows the expected quadratic behavior of the deformation
energy as a function of $u_0$, Eq. (\ref{deformation}), 
for a shear deformation in the small $u_0 $ limit. 
In the one and two-dimensional compressions ({\it i})-({\it iv}), however, 
Eq. (\ref{deformation}) leads to the addition of a non physical linear term in
the dependence of the energies $f_{dx},f_{dy},f_{dxy}$ on $u_{0}$ as can be
seen in Fig. 2. In the absence of deformation, the average electronic
density is equal to that of the positive background. This neutrality removes
the divergence of $H\left( \mathbf{G}\right) $ at $\mathbf{G}=0$ in the
Hartree-Fock energy of Eq. (\ref{energiehf}). When the electron lattice is
dilated (but not the positive background), the electronic density no longer
matches the density of the positive background and there is a restoring force
that arises from this density imbalance. It
is easy to show, assuming a density of the form of Eq. (\ref{m13}), that no
linear term in $u_{0}$ arises when the interaction with the positive
background is properly taken into account, and that the interaction with the
background does {\em not} give rise to higher order terms in $u_{0}$. Our
Hartree-Fock procedure requires that the electronic and background densities
be the same even for $u_{0}\neq 0$, which has the immediate consequence that 
we cannot directly compute the deformation energy using Eq. (\ref{deformation}). 
For all but the shear deformation, it is thus necessary for us to substract 
the linear term in Eq. (\ref{deformation}) and to add by hand the long-wavelength Coulomb
contribution of Eq. (\ref{defo}) in order to get the correct elastic
constants. Fig. 2 shows that a quadratic behavior for the deformation energy
is recovered when the linear term is substracted. Note that the deformation
energy in Fig. 2 does not contain the long-wavelength Coulomb contribution
of Eq. (\ref{defo}), so that it can be either positive or negative. Our definitions
in Eqs. (\ref{m2}, \ref{m3}, \ref{m4p}) of the elastic coefficients are not
affected by this procedure of removing the linear term in $u_{0}$ since they
involve the second derivative of the energy with respect to $u_{0}$.

\begin{figure}[tbph]
\includegraphics[scale=1]{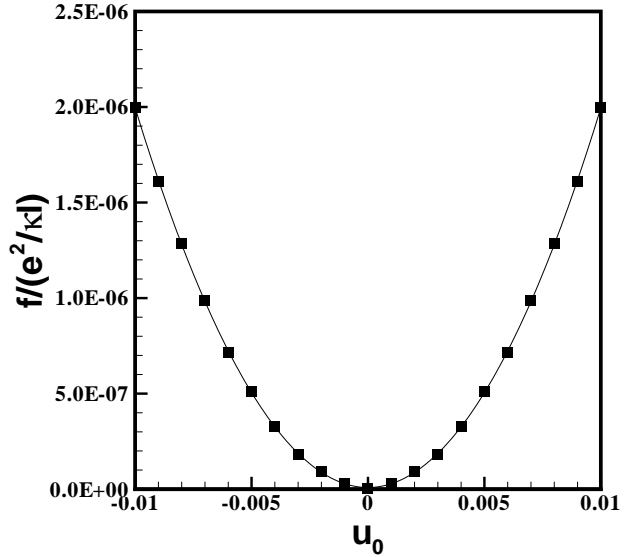}
\caption{Deformation energy as a function of $u_{0}$ for a shear
deformation: $u\left( \mathbf{r}\right) =u_{0}y\widehat{\mathbf{x}}$ in a
triangular Wigner crystal in Landau level $N=0$ with filling factor $\protect%
\nu =0.15$. The square symbols are the HFA result while the solid line is a
polynomial fit of order $2$ (the linear term is negligible). }
\end{figure}

\begin{figure}[tbph]
\includegraphics[scale=1]{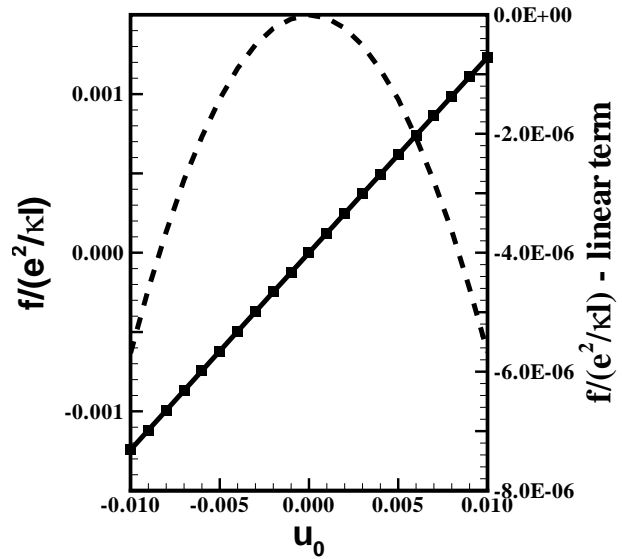}
\caption{Deformation energy as a function of $u_{0}$ for a one-dimensional
dilatation of a triangular Wigner crystal in Landau level $N=0$ with filling
factor $\protect\nu =0.15$. The square symbols are the HFA result (left
axis) while the solid line is a polynomial fit. The dashed line (right axis)
is the deformation energy with the linear term removed. }
\label{fig2a}
\end{figure}

It is instructive at this point to note that,
for a triangular Wigner crystal of classical electrons, the
calculation of Bonsall and Maradudin\cite{bonsall} gives the following expression
for the quantity $A(\nu)$:
\begin{equation}
A\left( \nu \right)=-0.782133\sqrt{\nu },
\end{equation}
and for the elastic coefficients: 
\begin{subequations}
\begin{eqnarray}
c_{12} &=&-0.10892\nu ^{3/2}\left( \frac{e^{2}}{\kappa \ell ^{3}}\right) ,
\label{m10} \\
c_{66} &=&0.015\,56\nu ^{3/2}\left( \frac{e^{2}}{\kappa \ell ^{3}}\right) .
\label{m11}
\end{eqnarray}
\end{subequations}
If we use Eqs. (\ref{m9}) (with the relation (\ref{triangulaire})) and take $%
c_{66}$ as given by Eq. (\ref{m11}), we find $c_{12}=-7c_{66}$, which is
consistent with Eq. (\ref{m10}).

\section{Dynamical matrix from the GRPA}

We now turn our attention to the calculation of the DM of electron crystals
in the GRPA method. In the strong magnetic field limit where the Hilbert space is restricted to
one Landau level only, the Hamiltonian of the system is given by: 
\begin{equation}
H=\sum_{\mathbf{k}}\sum_{\alpha ,\beta }u_{\alpha }\left( -\mathbf{k}\right)
D_{\alpha ,\beta }\left( \mathbf{k}\right) u_{\beta }\left( \mathbf{k}%
\right) ,
\end{equation}%
where $\alpha ,\beta =x,y$ and $\mathbf{k}$ is a vector restricted to the
first Brillouin zone of the crystal.
If we define the Matsubara displacement Green's function by: 
\begin{equation}
G_{\alpha ,\beta }\left( \mathbf{k},\tau \right) =-\left\langle 
T_{\tau }u_{\alpha }\left( \mathbf{k},\tau \right) 
u_{\beta }\left( -\mathbf{k},0\right) \right\rangle ,
\end{equation}%
we find, using $\hslash \frac{\partial }{\partial \tau }\left( \ldots \right) =\left[
H,\left( \ldots \right) \right] $ and the commutation relation $\left[ u_{x}(%
\mathbf{k}),u_{y}(\mathbf{k}^{\prime })\right] =i\ell ^{2}\delta _{\mathbf{k},\mathbf{-k}^{\prime }},$ that this Green's function is related to the
dynamical matrix by:
\begin{align}
G_{\alpha ,\beta }\left( \mathbf{k},i\Omega _{n}\right) & =\frac{-\ell ^{4}}{%
\hslash \left[ \Omega _{n}^{2}+\omega _{mp}^{2}\left( \mathbf{k}\right) %
\right] }  \label{m20} \\
& \times \left( 
\begin{array}{cc}
D_{y,y}\left( \mathbf{k}\right) & -\frac{\hslash \Omega _{n}}{\ell ^{2}}%
-D_{x,y}\left( \mathbf{k}\right) \\ 
\frac{\hslash \Omega _{n}}{\ell ^{2}}-D_{y,x}\left( \mathbf{k}\right) & 
D_{x,x}\left( \mathbf{k}\right)%
\end{array}%
\right) _{\alpha \beta },  \notag
\end{align}%
where $\Omega _{n}=2\pi n/T$ is a bosonic Matsubara frequency and 
\begin{equation}
\omega _{mp}\left( \mathbf{k}\right) =\frac{\ell ^{2}}{\hslash }\sqrt{\det %
\left[ D\left( \mathbf{k}\right) \right] }
\end{equation}%
is the magnetophonon dispersion relation.

We now define the following density Green's function: 
\begin{equation}
\chi _{\mathbf{G,G}^{\prime }}^{\left( \rho ,\rho \right) }\left( \mathbf{k}%
,\tau \right) =-N_{\varphi }\left\langle T_{\tau }\widetilde{%
\rho }\left( \mathbf{k}+\mathbf{G},\tau \right) \widetilde{\rho }
\left(-\mathbf{k}-\mathbf{G}^{\prime },0\right) \right\rangle ,  \label{m24}
\end{equation}%
where $\widetilde{\rho }=\rho -\left\langle \rho \right\rangle .$ In the
Generalised Random-Phase Approximation (GRPA), this Green's function is
found by solving the set of equations\cite{cotemethode}:
\begin{equation}
\sum_{\mathbf{G}^{\prime \prime }}\left[ i\Omega _{n}\delta _{\mathbf{G},%
\mathbf{G}^{\prime }}-M_{\mathbf{G},\mathbf{G}^{\prime \prime }}\left( 
\mathbf{k}\right) \right] \chi _{\mathbf{G}^{\prime \prime },\mathbf{G}%
^{\prime }}^{\left( \rho ,\rho \right) }\left( \mathbf{k},i\omega_{n}\right) 
=B_{\mathbf{G,G}^{\prime }}( \mathbf{q}),
\label{m21}
\end{equation}%
with the definitions:
\begin{align}
M_{\mathbf{G,G}^{\prime }}\left( \mathbf{k}\right) 
& =-2i\left( \frac{e^{2}}{\hslash \kappa \ell }\right) \left\langle 
\rho\left( \mathbf{G-G}^{\prime }\right) \right\rangle 
\nonumber\\
&\times
\sin \left[ \widehat{\mathbf{z}}\cdot \frac{\left( \mathbf{k}+\mathbf{G}\right) 
\times \left(\mathbf{k}+\mathbf{G}^{\prime }\right) \ell ^{2}}{2}\right]  \notag \\
& \times \big[ H\left( \mathbf{G}-\mathbf{G}^{\prime }\right) -X\left( 
\mathbf{G-G}^{\prime }\right) 
\nonumber\\
&-H\left( \mathbf{k}+\mathbf{G}^{\prime
}\right) +X\left( \mathbf{k}+\mathbf{G}^{\prime }\right) \big] , 
\end{align}
and:
\begin{align}
B_{\mathbf{G,G}^{\prime }}\left( \mathbf{k}\right) & = 
2i\left\langle\rho \left( \mathbf{G-G}^{\prime }\right) \right\rangle
\nonumber\\
&\times\sin \left[ \widehat{\mathbf{z}}
\cdot \frac{\left( \mathbf{q}+\mathbf{G}\right) \times \left( 
\mathbf{q}+\mathbf{G}^{\prime}\right) \ell ^2}{2}\right] .
\end{align}
Diagonalizing the matrix $M\left( \mathbf{k}\right) $ and making the
analytic continuation $i\Omega _{n}\rightarrow \omega +i\delta ,$ we can
write $\chi _{\mathbf{G,G}^{\prime }}^{\left( \rho ,\rho \right) }\left( 
\mathbf{k},\omega \right) $ in the form:
\begin{equation}
\chi _{\mathbf{G,G}^{\prime }}^{\left( \rho ,\rho \right) }\left( \mathbf{k}%
,\omega \right) =\sum_{i}\frac{W_{\mathbf{G,G}^{\prime }}^{(i)}\left( 
\mathbf{k}\right) }{\omega +i\delta -\widetilde{\omega }_{i}\left( \mathbf{k}%
\right) }.  \label{m25}
\end{equation}%
At small $\mathbf{k,}$ the pole $\widetilde{\omega }_{i}\left( \mathbf{k}
\right) $ with the biggest weight $W_{\mathbf{G,G}}^{(i)}\left( \mathbf{k}
\right) $ gives the GRPA\ magnetophon mode. We define this pole as $
\widetilde{\omega }_{GRPA}\left( \mathbf{k}\right) $ and the corresponding
weigth as $W_{\mathbf{G,G}}^{(GRPA)}\left( \mathbf{k}\right)$.

We now relate the displacement Green's function to the density Green's function
using Eq. (\ref{m7}). This last equation, coupled to Eq. (\ref{m24}), gives the following relation
between the density and displacement response functions (here $F(\mathbf{k})$
is the function defined in Eq. (\ref{nphi2}), where for simplicity we now
drop the Landau level index $N$): 
\begin{eqnarray}
\chi _{\mathbf{G,G}^{\prime }}^{\left( \rho ,\rho \right) }
\left( \mathbf{k},\omega \right) &=&\nu \frac{h(\mathbf{k+G})h(\mathbf{k+G}^{\prime })}{F(%
\mathbf{k+G})F(\mathbf{k+G}^{\prime })}  \label{r26} \\
&\times& \sum_{\alpha ,\beta }\left( k_{\alpha }+G_{\alpha }\right)
G_{\alpha ,\beta }\left( \mathbf{k},\omega \right) \left( k_{\beta
}+G_{\beta }^{\prime }\right) .  \notag
\end{eqnarray}%
In deriving Eq. (\ref{r26}), we have assumed that $\mathbf{q}\cdot \mathbf{u}
\left( \mathbf{R}\right) <<1$, so that a density fluctuation can be linearly
related to the displacement $\mathbf{u}\left( \mathbf{q}\right) $ by Eq. (%
\ref{m7}). This is equivalent to assuming that the crystal can be described
in the harmonic approximation so that only a knowledge of the dynamical
matrix is necessary. To get Eq. (\ref{r26}), we have also assumed that $%
h\left( \mathbf{r}\right) =h\left( -\mathbf{r}\right) $ so that $h\left( 
\mathbf{q}\right) \mathbf{\ }$is real. We can now use Eq. (\ref{m20}) and
the symmetry relation $D_{\alpha ,\beta }\left( \mathbf{k}\right) =D_{\beta
,\alpha }\left( \mathbf{k}\right) $ to relate the density response function
to the dynamical matrix:
\begin{eqnarray}
\chi _{\mathbf{G,G}^{\prime }}^{\left( \rho ,\rho \right) }\left( \mathbf{k},\omega \right) 
&=&\frac{\nu\ell^4}{\hbar}
\frac{\Big[ \Gamma _{1}({\bf k}) +\frac{i\hslash \omega }{\ell^2}\Gamma _{2}\left( \mathbf{k}\right) \Big]}
{\left[ \left( \omega+i\delta \right) ^{2}-\omega _{mp}^{2}\left( \mathbf{k}\right) \right] } 
\notag\\
&\times& \frac{h(\mathbf{k+G})h(\mathbf{k+G}^{\prime })}{F(\mathbf{k+G})F(%
\mathbf{k+G}^{\prime })},  
\end{eqnarray}
where we defined: 
\begin{subequations}
\begin{align}
&\Gamma _{1}\left( \mathbf{k}\right) =-\widehat{\mathbf{z}}\cdot \left[
\left( \mathbf{k}+\mathbf{G}\right) \times D\left( \mathbf{k}\right) \times
\left( \mathbf{k}+\mathbf{G}^{\prime }\right) \right] \cdot \widehat{\mathbf{z}},
\\
&\Gamma _{2}\left( \mathbf{k}\right) =\widehat{\mathbf{z}}\cdot \left[ \left( 
\mathbf{k}+\mathbf{G}\right) \times \left( \mathbf{k}+\mathbf{G}^{\prime
}\right) \right] .
\end{align}
\end{subequations}
For $\omega $ close to the magnetophonon resonance, we can write:
\begin{equation}
\chi _{\mathbf{G,G}^{\prime }}^{\left( \rho ,\rho \right) }\left( \mathbf{k}%
,\omega \right) \approx \frac{\nu \ell ^{4}}{\hslash }\frac{Z\left( \mathbf{k%
}\right) }{\omega +i\delta -\omega _{mp}\left( \mathbf{k}\right) }\frac{h(%
\mathbf{k+G})h(\mathbf{k+G}^{\prime })}{F(\mathbf{k+G})F(\mathbf{k+G}%
^{\prime })},  \label{m26}
\end{equation}%
where we defined the quantity:
\begin{equation}
Z\left( \mathbf{k}\right) =\frac{\Gamma _{1}\left( \mathbf{k}\right) }{%
2\omega _{mp}\left( \mathbf{k}\right) }+i\frac{\hslash \Gamma _{2}\left( 
\mathbf{k}\right) }{2\ell ^{2}}.
\end{equation}%
Then, equating Eq. (\ref{m26}) with Eq. (\ref{m25}) for $\omega $ close to 
$\omega_{GRPA}\left( \mathbf{k}\right) $, we obtain:
\begin{align}
\frac{\nu \ell ^{4}}{\hslash }
\frac{Z\left( \mathbf{k}\right) }{\omega+i\delta -\omega _{mp}\left( \mathbf{k}\right) }
&\frac{h(\mathbf{k+G})h(\mathbf{k+G}^{\prime })}{F(\mathbf{k+G})F(\mathbf{k+G}^{\prime })}  =
\notag \\
&\frac{W_{\mathbf{G,G}^{\prime }}^{(GRPA)}\left( \mathbf{k}\right) }{%
\omega +i\delta -\widetilde{\omega }_{GRPA}\left( \mathbf{k}\right) }.
\end{align}
Because $\omega _{mp}\left( \mathbf{k}\right) $ must be equal to $\widetilde{%
\omega }_{GRPA}\left( \mathbf{k}\right) $, we can finally write%
\begin{equation}
\frac{\nu \ell ^{4}}{\hslash }Z\left( \mathbf{k}\right) \frac{h(\mathbf{k+G}%
)h(\mathbf{k+G}^{\prime })}{F(\mathbf{k+G})F(\mathbf{k+G}^{\prime })}=W_{%
\mathbf{G,G}^{\prime }}^{(GRPA)}\left( \mathbf{k}\right) ,
\end{equation}%
or, taking the real and imaginary parts of this equation (we remind the
reader that both functions $h(\mathbf{k})$ and $F(\mathbf{k})$ are real): 
\begin{subequations}
\begin{eqnarray}
\Re \left[ W_{\mathbf{G,G}^{\prime }}^{(GRPA)}\left( \mathbf{k}\right) %
\right] &=&\frac{\nu }{2}\left[ \frac{h\left( \mathbf{k}+\mathbf{G}\right)
h\left( \mathbf{k}+\mathbf{G}^{\prime }\right) }{F\left( \mathbf{k}+\mathbf{G%
}\right) F\left( \mathbf{k}+\mathbf{G}^{\prime }\right) }\right]  \notag \\
&&\times \frac{\Gamma _{1}\left( \mathbf{k}\right) \ell ^{4}}{\hslash \omega
_{mp}\left( \mathbf{k}\right) }, \\
\Im \left[ W_{\mathbf{G,G}^{\prime }}^{(GRPA)}\left( \mathbf{k}\right) %
\right] &=&\frac{\nu }{2}\left[ \frac{h\left( \mathbf{k}+\mathbf{G}\right)
h\left( \mathbf{k}+\mathbf{G}^{\prime }\right) }{F\left( \mathbf{k}+\mathbf{G%
}\right) F\left( \mathbf{k}+\mathbf{G}^{\prime }\right) }\right]  \notag \\
&&\times \Gamma _{2}\left( \mathbf{k}\right) \ell ^{2}.
\end{eqnarray}%
We can get rid of the unknown form factors $h\left( \mathbf{k}+\mathbf{G}\right)$ 
if we work with the ratio of the imaginary and real parts of the
weights. We thus define:
\end{subequations}
\begin{eqnarray}
\Gamma _{\mathbf{G,G}^{\prime }}\left( \mathbf{k}\right) &\equiv &\frac{\Re %
\left[ W_{\mathbf{G,G}^{\prime }}^{(GRPA)}\left( \mathbf{k}\right) \right] }{%
\Im \left[ W_{\mathbf{G,G}^{\prime }}^{(GRPA)}\left( \mathbf{k}\right) %
\right] },  \label{m27} 
\\
&=&\frac{-\ell ^{2}}{\hslash \omega _{mp}\left( \mathbf{k}\right) }
\frac{\left( \mathbf{k}+\mathbf{G}\right) \times D\left( \mathbf{k}\right) \times
\left( \mathbf{k}+\mathbf{G}^{\prime }\right) }{\left( \mathbf{k}+\mathbf{G}%
\right) \times \left( \mathbf{k}+\mathbf{G}^{\prime }\right) }.  \notag
\end{eqnarray}%
A careful examination shows that, because $\omega _{mp}\left( \mathbf{k}%
\right) $ is given by the determinant of the dynamical matrix $D\left( 
\mathbf{k}\right) $, the quantity $\Gamma _{1}\left( \mathbf{k}\right)
/\omega _{mp}\left( \mathbf{k}\right) $ is unchanged if all the components
of the dynamical matrix are multiplied by some constant. Eq.~(\ref{m27}) is
thus indeterminate. To avoid this problem, we replace $\omega _{mp}\left( 
\mathbf{k}\right) $ by $\widetilde{\omega }_{GRPA}\left( \mathbf{k}\right) $
in Eq. (\ref{m27}). Our final result is thus:%
\begin{equation}
\Gamma _{\mathbf{G,G}^{\prime }}\left( \mathbf{k}\right) =\frac{-\ell ^{2}}{%
\hslash \widetilde{\omega }_{GRPA}\left( \mathbf{k}\right) }\frac{\left( 
\mathbf{k}+\mathbf{G}\right) \times D\left( \mathbf{k}\right) \times \left( 
\mathbf{k}+\mathbf{G}^{\prime }\right) }{\left( \mathbf{k}+\mathbf{G}\right)
\times \left( \mathbf{k}+\mathbf{G}^{\prime }\right) }.  \label{m29}
\end{equation}

Because $D_{x,y}\left( \mathbf{k}\right) =D_{y,x}\left( \mathbf{k}\right) $,
we need to choose three pairs of vectors $\left( \mathbf{G,G}^{\prime
}\right) $ to get the components of the dynamical matrix. To be valid, the
dynamical matrix obtained in this way must satisfy the equation:%
\begin{equation}
\widetilde{\omega }_{GRPA}\left( \mathbf{k}\right) =\frac{\ell ^{2}}{\hslash 
}\sqrt{\det \left[ D\left( \mathbf{k}\right) \right] }.  \label{m33}
\end{equation}
Eq. (\ref{m33}) provides a check on the validity of our calculation.

\section{Numerical results for the Wigner crystal}

\begin{figure}[tbph]
\includegraphics[scale=0.8]{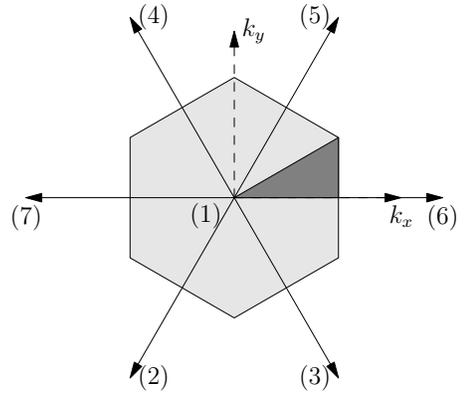}
\caption{First Brillouin zone of the triangular lattice with the irreducible
Brillouin zone shown as the dark area. The arrows represent the reciprocal lattice vectors
on the second shell while (1) corresponds to the vector $\mathbf{G}=0$. }
\label{fig3}
\end{figure}

In this Section, we illustrate the application of our method by computing
the Lam\'{e} coefficients for the triangular Wigner crystal in Landau levels 
$N=0$ and $N=2$. Fig. \ref{fig3} shows the first two shells of reciprocal
lattice vectors of the triangular lattice with $\mathbf{G}_{1}=\left(
0,0\right) .$ We take the vectors $\mathbf{G}$ and $\mathbf{G}^{\prime }$ in
Eq. (\ref{m29}) on these first two shells. Not all combinations of vectors
satisfy Eq. (\ref{m33}). By experimentation, we found that with a
combination of the form $\left[ \left( \mathbf{G},0\right),\left( \mathbf{G}
^{\prime },0\right) ,\left( \mathbf{G},\mathbf{G}^{\prime }\right) \right] $
with $\mathbf{G},\mathbf{G}^{\prime }\neq 0,$ this equation is satisfied in
the irreducible Brillouin zone shown in Fig. 3 to better than $0.05\%$ for 
$k\ell \lesssim 0.3$. We will thus stick to this type of combination for the
rest of this paper. 

In the small-wavevector limit, the dynamical matrix of the Wigner crystal
with a triangular lattice structure is given by Eq. (\ref{m32}). Using Eq. (%
\ref{m29}), we can extract the elastic coefficients by fitting $D_{\alpha,\beta }\left( \mathbf{k}\right)$ 
along the path $k_y=0$ where:
\begin{figure}[tbph]
\includegraphics[scale=1]{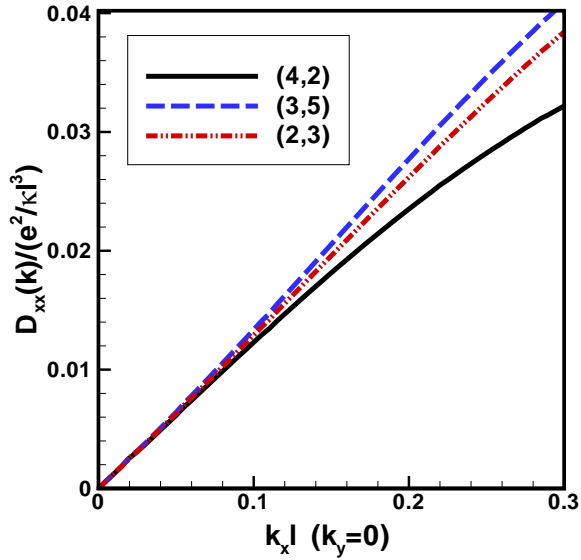}
\caption{(Color online) Component $D_{xx}(\mathbf{k})$ of the dynamical
matrix along $k_{x}$ computed for differents couples of vectors $(\mathbf{G},%
\mathbf{G}^{\prime })$. The numbers in the legend refer to the numerotation
of the reciprocal lattice vectors shown in Fig. 3. }
\label{fig4}
\end{figure}

\begin{subequations}
\begin{align}
&D_{x,x}\left( \mathbf{k}\right) = \left( \frac{e^{2}}{\kappa \ell ^{3}}%
\right) \nu k_{x}\ell +\frac{2\pi }{\nu }\left( c_{12}+2c_{66}\right)
k_{x}^{2}\ell ^{2},  
\label{m35} \\
&D_{x,y}\left( \mathbf{k}\right)  = 0,  
\label{m36} \\
&D_{y,y}\left( \mathbf{k}\right)  = \frac{2\pi }{\nu }c_{66}k_{x}^{2}\ell^{2},  
\label{m37}
\end{align}
\end{subequations}
where $c_{12}$ and $c_{66}$ are expressed in units of $e^{2}/\kappa \ell ^{3}$.

Fig. \ref{fig4} illustrates one limitation of our method: the GRPA\ dynamical matrix
is very much dependent on the choice of the couple ($\mathbf{G},\mathbf{G}%
^{\prime }$). Different choices give the same dynamical matrices $D_{\alpha
,\beta }\left( \mathbf{k}\right) $ only in the small wavector limit $k\ell
\lesssim 0.1$ as shown in Fig. 4 and, in this limit, the dynamical matrix
element $D_{x,x}\left( \mathbf{k}\right) $ is almost entirely dominated by
the long-range Coulomb term (the first term in Eq. (\ref{m35})). It follows
that different choices of ($\mathbf{G},\mathbf{G}^{\prime }$) lead to quite
different values of the elastic coefficients $c_{12}$ even though Eq. (\ref%
{m33}) is satisfied. The coefficient $c_{66}$ obtained from 
$D_{y,y}\left( \mathbf{k}\right)$, however, is not affected by the long-range
Coulomb interaction and appears to be independent of the choice of ($\mathbf{G},%
\mathbf{G}^{\prime }$). Note that the dynamical matrix given by Eq. (\ref%
{m29}) does not have the correct transformation symmetries of the triangular
lattice. In cases where $D_{\alpha ,\beta }\left( \mathbf{k}\right) $ is needed in all
the Brillouin zone, it becomes necessary to compute 
$D_{\alpha, \beta }\left( \mathbf{k}\right)$ 
in the irreducible Brillouin zone and obtain $D_{\alpha,\beta }\left( \mathbf{k}\right) $ 
in the rest of the Brillouin zone by
symmetry.

\begin{figure}[tbph]
\includegraphics[scale=1]{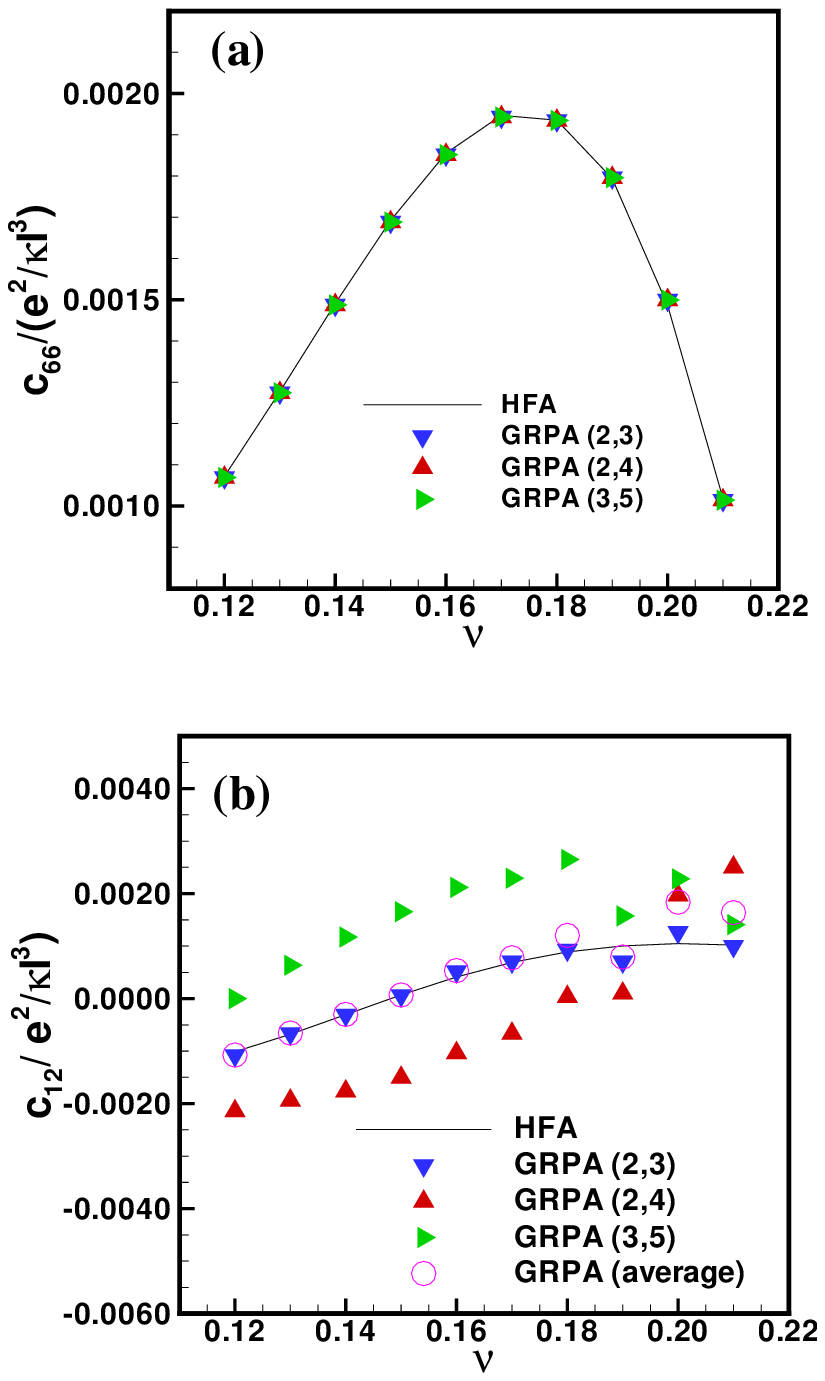}
\caption{(Color online) Elastic coefficients (a) $c_{66}$ and (b) $c_{12}$ of the
triangular Wigner crystal for Landau level $N=2$ computed using the
different approximations listed in the legend. For the GRPA, the
coefficients are computed using $3$ differents couples of reciprocal lattice
vectors. The numbers in the legend correspond to the numeration of the
vectors given in Fig. 3. The empty circles give an average of the 3 GRPA\
results. For $c_{66}$ the different symbols are superimposed.}
\label{fig5}
\end{figure}

\begin{figure}[tbph]
\includegraphics[scale=1]{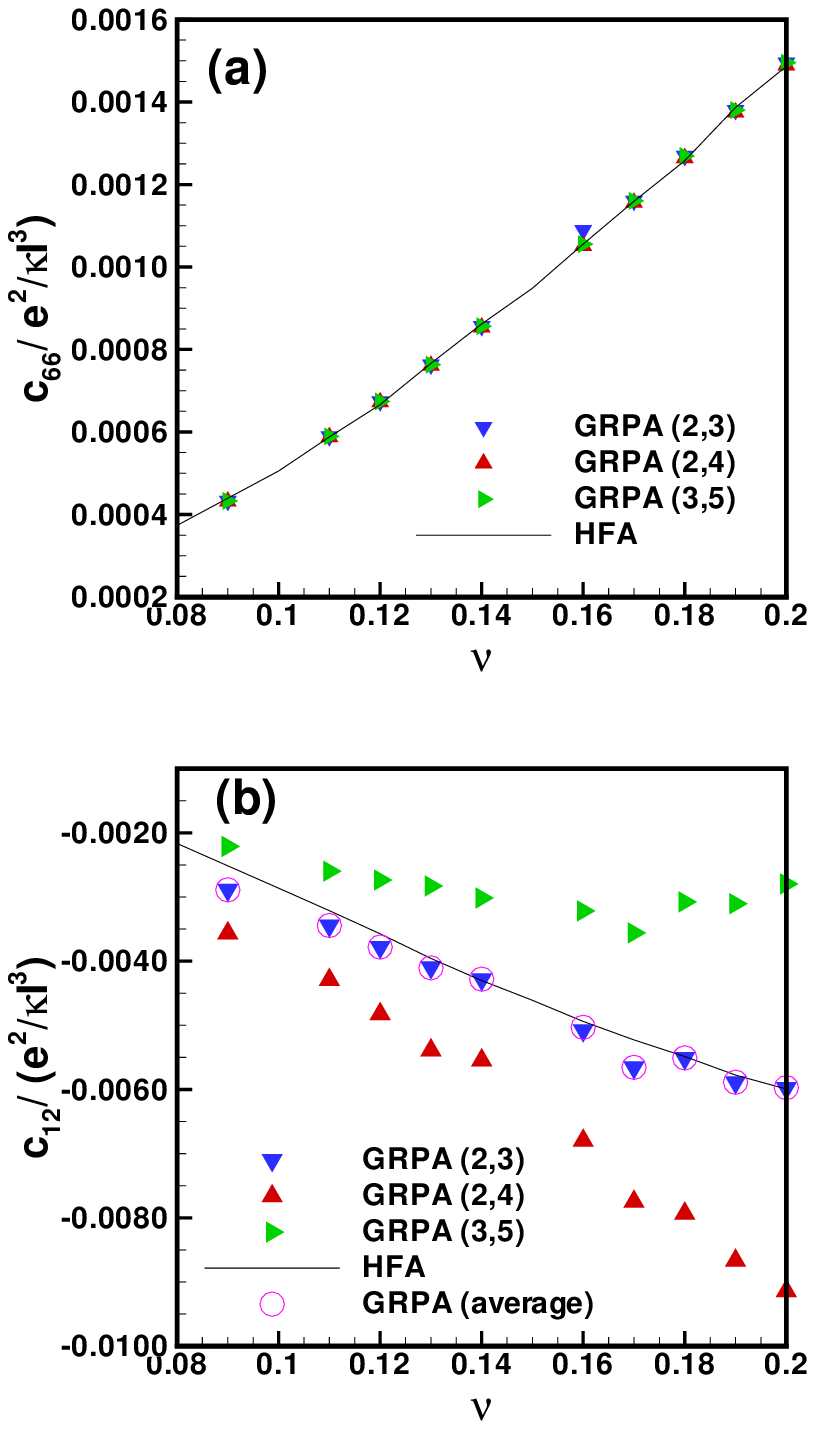}
\caption{(Color online) Elastic coefficients (a) $c_{66}$ and (b) $c_{12}$ of the
triangular Wigner crystal in Landau level $N=0$ computed in the different
approximations indicated in the legend. For the GRPA, the coefficients are
computed using $3$ differents couples of reciprocal lattice vectors. The
numbers in the legend correspond to the numeration of the vectors given in
Fig. 3. The empty circles give the average of the 3 GRPA\ results. For $%
c_{66},$ the different symbols are superimposed.}
\label{fig6}
\end{figure}

To give an idea of the variability of the numerical results with $(\mathbf{G},\mathbf{G}^{\prime })$,
we show in Fig. \ref{fig5} (for $N=2$) and Fig. \ref{fig6} (for $N=0$) the coefficients 
$c_{12}$ and $c_{66}$ extracted from the GRPA dynamical matric of the
triangular Wigner crystal for different couples of vectors 
$(\mathbf{G},\mathbf{G}^{\prime })$.
These coefficients are compared with those computed using the HFA described
in Sec. III. We show the HFA results by a full line in Figs. 5,6. For both 
$N=0$ and $N=2$, we find that the Hartree-Fock results for the coefficient 
$c_{66}$ are extremely well reproduced by the GRPA\ method and, as we said
above, do not depend on the choice of ($\mathbf{G},\mathbf{G}^{\prime }$).
This is what we expect since the GRPA\ is the linear response of the crystal
about the HFA ground state so that the coefficients $c_{ij}$ obtained from
the two methods should be roughly equal, taking into account the various
approximations made in deriving the GRPA\ dynamical matrix. The coefficient 
$c_{66}$ is easy to obtain, in view of Eq. (\ref{m37}), since it is given by
a one-parameter fit of the $D_{y,y}\left( k_{x},k_{y}=0\right) $ curve.
The elastic coefficient $c_{12}$ (which is related to the bulk modulus) is, on the other hand, much
more difficult to obtain from Eq. (\ref{m35}). Indeed, this elastic coefficient turns 
out to be very sensitive to how accurately the long-wavelength limit $\nu k_{x}\ell $ 
of $D_{x,x}\left( \mathbf{k}\right) $ in Eq. (\ref{m35}) is obtained by the GRPA numerical
calculation. (We here note that the GRPA dynamical matrix {\em does} contain the long-range Coulomb
interaction discussed in Sec. II. The latter does not have to be added by hand as
was the case for the elastic coefficients computed in the HFA.) As we see
in Figs. 5(b) and 6(b), $c_{12}$ is also very sensitive to the choice of
the vectors ($\mathbf{G},\mathbf{G}^{\prime }$) with one particular choice
(2,3) reproducing the HFA results almost exactly. The other two choices give
very different values for $c_{12}$. In the absence of any criteria to
choose ($\mathbf{G},\mathbf{G}^{\prime }$) a priori, we would say that the
GRPA\ dynamical matrix cannot be used to make \textit{quantitative}
predictions. The \textit{qualitative} behaviour of the GRPA\ elastic
coefficient $c_{12}$ is consistent with that of $c_{12}$ computed in the
HFA.

If we exclude the domain $\nu \geq 0.19$ where 
our numerical results become noisy, we find that the average of the GRPA results for the three couples
of ($\mathbf{G},\mathbf{G}^{\prime }$), as shown in Figs. 5(b) and 6(b), are
in very good agreement with the HF calculation. In the absence of any
criteria to choose the best couple ($\mathbf{G},\mathbf{G}^{\prime }$), this
averaging procedure must be used to get qualitatively and quantitatively
reliable results for the GRPA dynamical matrix. For $\nu \geq 0.19$, the
crystal softens and the quantum fluctuations in $\mathbf{u}$ are important.
There is a transition\cite{cote} into a bubble state with 2 electrons per
unit cell at approximately $\nu =0.22$. We do not expect the assumptions
underlying our method to be valid in this region.

\section{Numerical results for the stripe crystal}

For the stripe crystal, the dynamical matrix is given by: 
\begin{subequations}
\begin{eqnarray}
D_{x,x}\left( \mathbf{k}\right) &=&\left( \frac{e^{2}}{\kappa \ell ^{3}}
\right) \frac{\nu }{k\ell }k_{x}^{2}\ell ^{2}+\frac{2\pi }{\nu }\big(
c_{11}k_{x}^{2}\ell ^{2}+c_{66}k_{y}^{2}\ell ^{2}
\notag\\
&+&Kk_{y}^{4}\ell^{4}\big) ,  
\\
D_{x,y}\left( \mathbf{k}\right) &=&\left( \frac{e^{2}}{\kappa \ell ^{3}}%
\right) \frac{\nu }{k\ell }k_{x}k_{y}\ell ^{2}+\frac{2\pi }{\nu }\left(
c_{12}+c_{66}\right) k_{x}k_{y}\ell ^{2},  
\nonumber\\ 
\\
D_{y,y}\left( \mathbf{k}\right) &=&\left( \frac{e^{2}}{\kappa \ell ^{3}}%
\right) \frac{\nu }{k\ell }k_{y}^{2}\ell ^{2}+\frac{2\pi }{\nu }\left(
c_{22}k_{y}^{2}\ell ^{2}+c_{66}k_{x}^{2}\ell ^{2}\right) ,\nonumber\\  
\end{eqnarray}
\label{zz}
\end{subequations}
and the elastic coefficients evaluated in the HFA\cite{erratum} for Landau
level $N=2$ are listed in Table \ref{table1}.

\begin{table}[tbp] \centering%
\begin{tabular}{|l|l|l|l|l|}
\hline
$\nu $ & $c_{11}\left( \times 10^{-2}\right) $ & $c_{12}\left( \times
10^{-1}\right) $ & $c_{22}\left( \times 10^{-2}\right) $ & $c_{66}\left(
\times 10^{-5}\right) $ \\ \hline
$0.42$ & $7.13$ & $-2.43$ & $-0.73$ & $4.35$ \\  \hline
$0.43$ & $5.91$ & $-2.47$ & $-1.15$ & $6.31$ \\  \hline
$0.44$ & $4.95$ & $-2.49$ & $-1.64$ & $7.08$ \\ \hline
$0.45$ & $4.21$ & $-2.50$ & $-2.15$ & $7.10$ \\ \hline
$0.46$ & $3.75$ & $-2.51$ & $-2.65$ & $6.21$ \\ \hline
\end{tabular}%
\caption{Elastic coefficients  $n_0^{-1}c_{i,j}$ in units of $e^2/\kappa\ell$ for the stripe
crystal at various filling factors and in Landau level $N=2$.}%
\label{table1}
\end{table}

\begin{figure}[tbph]
\includegraphics[scale=0.75]{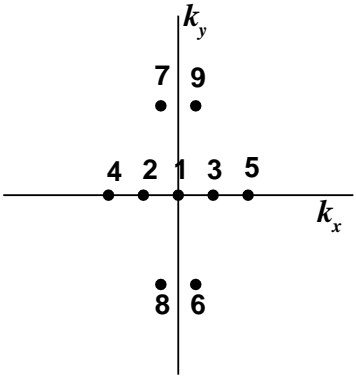}
\caption{The first four shells of reciprocal lattice vectors of the
anisotropic stripe cristal. }
\end{figure}

The first 4 shells of reciprocal lattice vectors of the stripe crystal are
represented in Fig. 7. From Eq. (\ref{m29}), the vectors ($\mathbf{G},
\mathbf{G}^{\prime }$) must not be parallel otherwise the denominator in
this equation vanishes. This forces us to use $\mathbf{G}$ in the second
shell and $\mathbf{G}^{\prime }$ in the fourth shell of reciprocal lattice
vectors to evaluate the DM in the GRPA. We show in Figs. 8-10 the elements $D_{xx},D_{xy},$ and $D_{yy}$ computed at filling factor $\nu =0.42$ (in
Landau level $N=2$) along different directions in $\mathbf{k}$-space
together with the corresponding DM in the HFA element obtained from Eqs. (\ref{zz})
with the coefficients of Table \ref{table1}. Similar results are obtained at other
filling factors. Notice that the bending coefficient $K\ $does not
contribute to any of these curves. For each curve, Eq. (\ref{m33}) is
perfectly satisfied and the coefficient $c_{66}$, which can be extracted
from the GRPA$\ $function $D_{yy}\left( k_{x},k_{y}=0\right)$, is in
excellent agreement with the HFA results given in Table \ref{table1}.

\begin{figure}[tbph]
\includegraphics[scale=1]{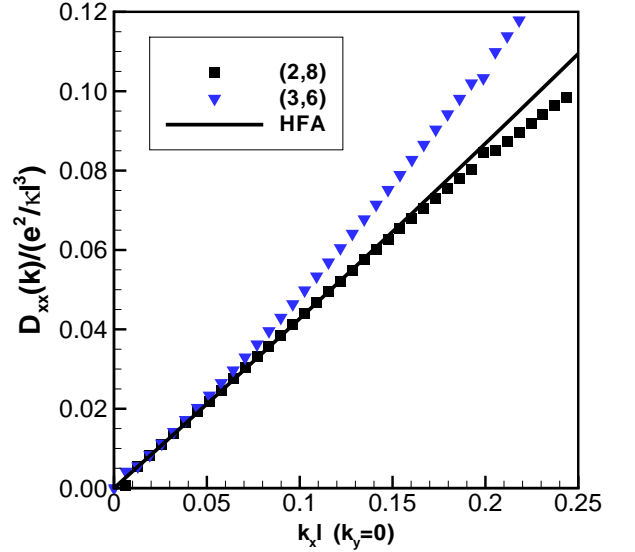}
\caption{(Color online) Component $D_{xx}(\mathbf{k})$ of the GRPA and HFA
dynamical matrices of the stripe crystal computed along the direction $%
k_{y}=0$ for partial filling factor $\protect\nu =0.42$ in Landau level $N=2$%
, computed using $2$ differents couples of reciprocal lattice vectors. The
numbers in the legend refer to numbering of the reciprocal lattice vectors
in Fig. 7.}
\end{figure}

\begin{figure}[tbph]
\includegraphics[scale=1]{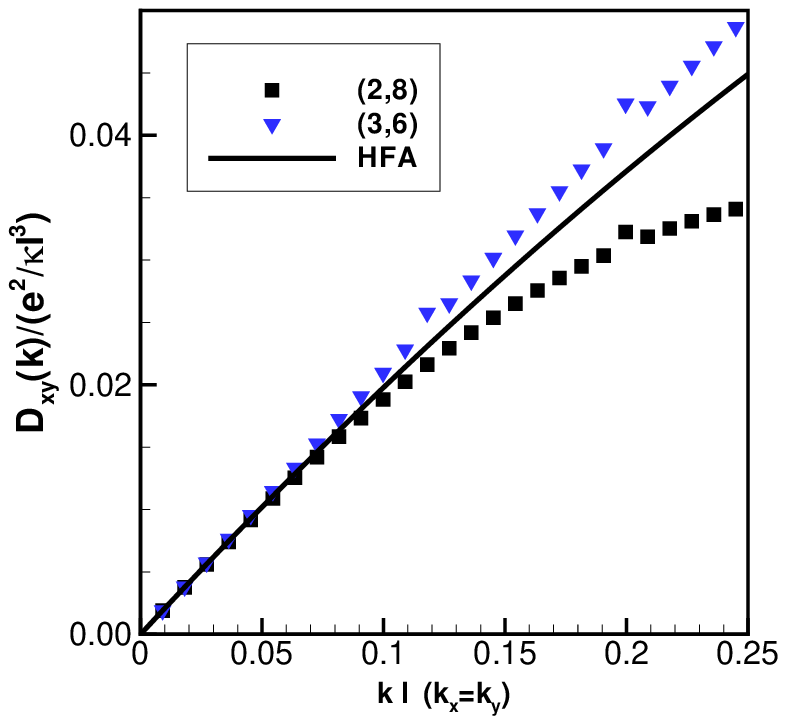}
\caption{(Color online) Component $D_{xy}(\mathbf{k})$ of the GRPA and HFA
dynamical matrices of the stripe crystal computed along the direction $%
k_{y}=k_{x}$ for partial filling factor $\protect\nu =0.42$ in Landau level $%
N=2$, computed using $2$ differents couples of reciprocal lattice vectors.
The numbers in the legend refer to numbering of the reciprocal lattice
vectors in Fig. 7.}
\end{figure}

\begin{figure}[tbph]
\includegraphics[scale=1]{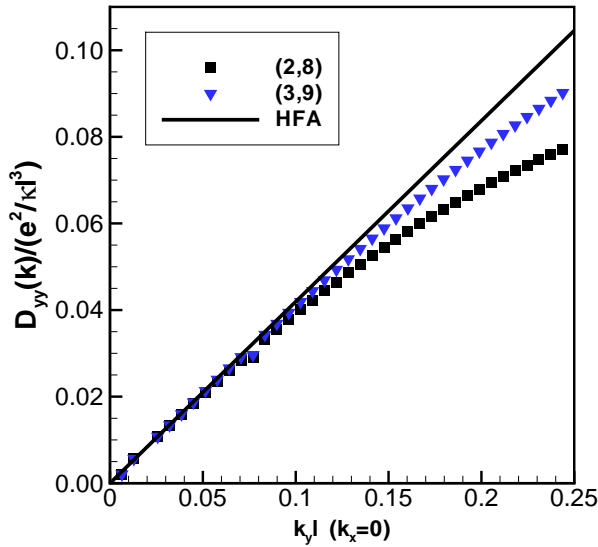}
\caption{(Color online) Component $D_{yy}(\mathbf{k})$ of the GRPA and HFA
dynamical matrices of the stripe crystal computed along the direction $%
k_{x}=0$ for partial filling factor $\protect\nu =0.42$ in Landau level $N=2$%
, computed using $2$ differents couples of reciprocal lattice vectors. The
numbers in the legend refer to numbering of the reciprocal lattice vectors
in Fig. 7.}
\end{figure}

For the GRPA, Figs. 8-10 show results for the couples ($\mathbf{G},\mathbf{G}^{\prime }$) 
that produce the maximum and minimum values of the DM element.
In all but the $D_{yy}(k_{x}=0,k_{y})$ case, the HFA curve lies between these
two results. For $D_{yy}(k_{x}=0,k_{y})$, one of the GRPA curves almost
coincides with the HFA result for $k_{y}\ell \lesssim 0.15.$ This is
reassuring for the validity of the GRPA method, but it also makes it
impossible for us to find what part of the difference between the GRPA
and HFA is numerical and what part is physical (i.e. due to anharmonicity
for example). We remark that, in the range $k\ell \lesssim 0.15$, the GRPA\
results are not numerically very different from the small $k\ell $ expansion
of the dynamical matrix given in Eqs. (\ref{zz}) with the HFA coefficients.
To the credit of our GRPA method, we add that the evolution of the different 
$D_{i,j}$'s with filling factor is consistent with that of the corresponding
elements calculated in the HFA as shown in Fig. 11.

We thus see that the evaluation of the elastic coefficients other than $c_{66}$ from the
GRPA results seems hazardous for the stripe crystal. The curvature of the
functions $D_{xx},D_{xy},$ and $D_{yy}$ in Figs. 8,9,10 is proportional to $%
c_{11},c_{12}+c_{66}$ and $c_{22}$ respectively. It is clear that the
elastic coefficients extracted from these $D_{i,j}$ are much bigger than
those obtained from the HFA (the curvature of the HFA\ function is barely
visible in the figures). These coefficients also show very strong variation
with the choice of ($\mathbf{G},\mathbf{G}^{\prime }$). An averaging of the
GRPA\ results for different couples ($\mathbf{G},\mathbf{G}^{\prime }$)
would give a result closer to the HFA but the improvement would not be as
dramatic as in the triangular lattice case. In fact, in the case of $D_{yy}$,
we find that averaging over different choices of reciprocal lattice vectors
does not bring any improvement to the numerical results.

\begin{figure}[tbh]
\includegraphics[scale=1]{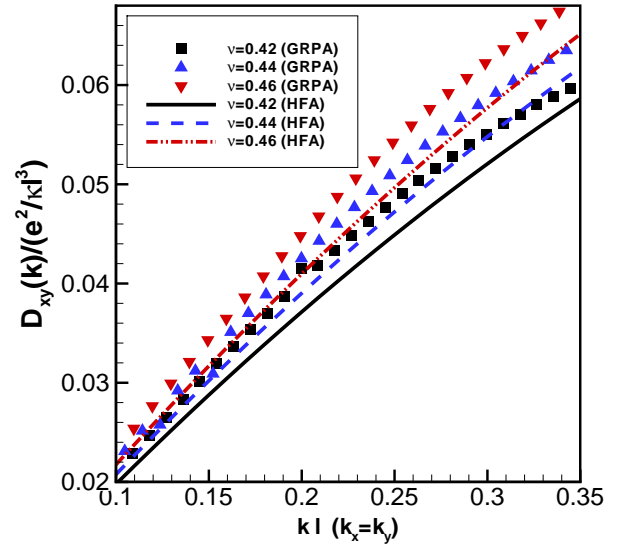}
\caption{(Color online) Component $D_{xy}(k)$ of the GRPA and HFA dynamical
matrices of the stripe crystal computed along the direction $k_{x}=k_{y}$
for different filling factors in Landau level $N=2$.}
\end{figure}

Finally, we remark that our GRPA results for $D_{xx}\left(
k_{x}=0,k_{y}\right)$ are dominated by a strong $k_{y}^{4}$ behavior
indicating that the bending term $K$ is absolutely essential in the elastic
description of the stripe crystal in Eqs. (\ref{zz}).

\section{Conclusion}

In conclusion, in this paper we have shown that is it possible to derive an effective
dynamical matrix for various crystal states of the 2DEG in a strong magnetic
field by computing the density response function in the GRPA. We have
compared the dynamical matrix obtained in this way with the one obtained from
standard elasticity theory with elastic coefficients computed in the HFA.
Our comparison was done for crystals with very different elastic properties,
namely a triangular Wigner crystal and stripe crystal.
Our motivation for deriving a dynamical matrix using the GRPA response consists in the fact
that the latter has the advantage of giving the dynamical matrix directly without
having to compute the elastic coefficients separately. Our comparison with
the Hartree-Fock results showed, however, that the GRPA method must be used
with care because of the variability of the results with the choice of
the couples $\left( \mathbf{G},\mathbf{G}^{\prime }\right)$. The shear
modulus $c_{66}$ computed in the GRPA agrees very well with the one computed
from the HFA, but the values of the other elastic coefficients $c_{ij}$ which are
affected by the long range Coulomb interaction depend very much on the
choice of the couples $\left( \mathbf{G},\mathbf{G}^{\prime }\right)$. In
some cases, as for a triangular Wigner crystal, an averaging procedure over
different couples $\left( \mathbf{G},\mathbf{G}^{\prime }\right) $ improves
the numerical accuracy of the method. In the long wavelength $k\ell\ll 1$ limit, however,
the GRPA dynamical matrix is a good approximation, both qualitatively and quantitatively,
and gives reasonable estimates for the elastic constants of the electronic solids
that are in agreement with the static Hartree-Fock calculations.

\begin{acknowledgments}
This work was supported by a research grant (for R. C\^{o}t\'{e}) from the
Natural Sciences and Engineering Research Council of Canada (NSERC). C. B.
Doiron acknowledges support from NSERC, the Fonds qu\'{e}b\'{e}cois de
recherche sur la nature et les technologies (FQRNT), the Swiss NSF and NCCR
Nanoscience. Computer time was provided by the R\'{e}seau qu\'{e}b\'{e}cois
de calcul haute performance (RQCHP).
\end{acknowledgments}


\begin{thebibliography}{99}
\bibitem{erratum} These coefficients were evaluated previously in Refs. %
\onlinecite{ettouhami1,ettouhami2}. We remark however that there was an
error in the evaluation of the coefficient $c_{12}$ in Ref. %
\onlinecite{ettouhami2} that we have corrected in the present paper. Our
previous calculation underestimated the value of $c_{12}$ by a factor of
approximately $3$. Note also that in the present paper the stripes are
aligned along the $y$ axis while they were aligned along $x$ in our previous
calculation so that $c_{11}\leftrightarrow c_{22}$.
\end{thebibliography}
\end{document}